\documentclass[aps,prd,showpacs,floatfix,twoside,%
twocolumn,nofootinbib,byrevtex]{revtex4}  

\usepackage{epsfig,bm,amsmath,amssymb}

\begin{document}

\preprint{nucl-th/0702077}


\title{\boldmath $J/\psi$ absorption by nucleons in the meson-exchange model}


\author{Yongseok Oh}%
\email{yoh@comp.tamu.edu}

\author{Wei Liu}%
\email{weiliu@comp.tamu.edu}

\author{C. M. Ko}%
\email{ko@comp.tamu.edu}

\affiliation{
Cyclotron Institute and Physics Department,
Texas A\&M University, College Station, TX 77843, U.S.A.}

\date{\today}


\begin{abstract}

We reinvestigate the $J/\psi$ dissociation processes induced by the
reactions with nucleons, $J/\psi + N \to \bar{D}^{(*)} + \Lambda_c$,
in the meson-exchange model. Main constraints used in this work are
vector-meson dominance and charm vector-current conservation. We
show that the cross section for $J/\psi + N \to \bar{D} + \Lambda_c$
can be larger than that for $J/\psi + N \to \bar{D}^{*} + \Lambda_c$
when these constraints are imposed. The dependence of the cross
sections on the coupling constants is analyzed in detail, and the
comparison with the recent quark-interchange model predictions is
also made.

\end{abstract}


\maketitle

\section{Introduction}

Since the suggestion of $J/\psi$ suppression as a signal for the
formation of the quark-gluon plasma in relativistic heavy ion
collisions \cite{MS86}, understanding the interactions of the
$J/\psi$ with other hadrons has been an important issue as $J/\psi$
dissociation by hadrons could also cause the suppression of the
produced $J/\psi$ \cite{FLP88-GH88-GGJ88-VPKH88}. Because the
$J/\psi$-hadron interactions cannot be directly accessed by present
experiments, the $J/\psi$-hadron cross sections have been estimated
through several assumptions and/or model calculations. Empirically,
the $J/\psi$-nucleon cross sections have been estimated by using the
data for $J/\psi$ photoproduction from the nucleon
\cite{HK98-RSZ00}, $J/\psi$ photoproduction from nuclei, and
$J/\psi$ production from the nucleon-nucleus collisions
\cite{KLNS97}. Since those data are scattered over a wide range of
energy and the estimation is model-dependent, the estimated values
for the $J/\psi$-nucleon cross sections range from $\sim 1$ mb to
$\sim 7$ mb. (See also Ref.~\cite{AT06} for a recent study on this subject.)

Theoretically, these cross sections have been estimated in various
ways including the perturbative QCD \cite{pqcd}, QCD sum-rule
approach \cite{qcdsr}, meson-exchange models
\cite{MM98a,Haglin00-STT01,LK00,OSL01,LKL02}, Regge theory approach
\cite{IL03-LC06}, quark models \cite{qm,HBBS07}, lattice QCD
\cite{YSHH06}, and other methods \cite{DNNR99-SV05b}. Despite of the
efforts to resolve the model-dependence of the cross sections, the
uncertainties in theoretical/model calculations for the
$J/\psi$-hadron cross sections are not yet clarified and the
predicted cross sections are model-dependent not only in the
magnitude but also in the energy dependence.

Recently, the $J/\psi$-nucleon dissociation cross sections have been
calculated in a quark-interchange model by Hilbert {\it et al.\/}
\cite{HBBS07}, and the results show a large difference from the
meson-exchange model predictions of Ref.~\cite{LKL02}. In
particular, the two models predict very different values for the
ratio of the cross sections,%
\footnote{The ratio $R_{D/D^*}$ depends on the energy. Here, we
compare the peak values of the two cross sections.} $R_{D/D^*} =
\sigma(J/\psi + N \to \bar{D} + \Lambda_c)/\sigma(J/\psi + N \to
\bar{D}^* + \Lambda_c)$, namely, $R_{D/D^*} > 50$ in
Ref.~\cite{HBBS07}, whereas $R_{D/D^*} < 0.5$ in the model of
Ref.~\cite{LKL02}. In this paper, in order to understand this
discrepancy, we re-examine and improve the meson-exchange model of
Ref.~\cite{LKL02} by using vector-meson dominance and charm
vector-current conservation to constrain the coupling constants and
form factors of this model. We will show that these constraints can
lead to $R_{D/D^*} > 1$ in the meson-exchange model. We will also
discuss the role driven by the tensor coupling terms of the
interactions of vector mesons.

This paper is organized as follows. In the next Section, we discuss
the coupling constants of the relevant effective Lagrangians for the
$J/\psi$-nucleon absorption processes. The $J/\psi$ coupling
constants are discussed in connection with vector-meson dominance,
which leads to the universality of the $J/\psi$ coupling. The role
of the conserved charm vector-current in determining coupling
constants and in constraining form factors are also explained.
Section~III contains the numerical results, and the dependence of
the cross sections on the coupling constants and form factors are
explored. We summarize in Sec.~IV.

\section{The model}

\begin{figure*}[t]
\centering \epsfig{file=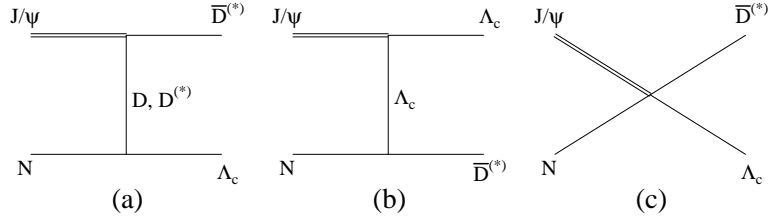, width=0.6\textwidth}
\caption{Diagrams for the reaction of $J/\psi + N \to \bar{D}^{(*)}
+ \Lambda_c$.} \label{fig:diag}
\end{figure*}

The diagrams which contribute to the $J/\psi$-nucleon dissociation
reactions are shown in Fig.~\ref{fig:diag}. To evaluate these
diagrams, we use the following effective Lagrangians:
\begin{eqnarray}
\mathcal{L}_{\psi DD} &=& -i g_{\psi DD}^{} \psi^\mu \left( D
\partial_\mu \bar{D} - \partial_\mu D \bar{D} \right),
\nonumber \\
\mathcal{L}_{\psi D^*D^*} &=& -i g_{\psi D^*D^*}^{} \left\{ \psi^\mu (
\partial_\mu D^{*\nu} \bar{D}_\nu^* - D^{*\nu} \partial_\mu
\bar{D}_\nu^* )
\right. \nonumber \\ && \mbox{}
+ (\partial_\mu \psi_\nu D^{*\nu} - \psi_\nu \partial_\mu D^{*\nu})
\bar{D}^{*\mu}
\nonumber \\ && \left. \mbox{}
+ D^{*\mu} ( \psi^\nu \partial_\mu \bar{D}_\nu^* - \partial_\mu \psi_\nu
\bar{D}^{*\nu} ) \right\},
\nonumber \\
\mathcal{L}_{\psi D^*D} &=& -g_{\psi D^*D}^{}
\varepsilon^{\mu\nu\alpha\beta} \partial_\mu \psi_\nu ( \partial_\alpha
D_\beta^* \bar{D} + D \partial_\alpha \bar{D}_\beta^*),
\nonumber \\
\mathcal{L}_{\psi\Lambda_c\Lambda_c} &=& g_{\psi\Lambda_c\Lambda_c}^{}
\overline{\Lambda}_c \left( \gamma_\mu \psi^\mu -
\frac{\kappa_{\psi\Lambda_c\Lambda_c}^{}}{2M_N}
\sigma_{\mu\nu} \partial^\nu \psi^\mu \right) \Lambda_c,
\nonumber \\
\mathcal{L}_{DN\Lambda_c} &=& -ig_{DN\Lambda_c}^{} \overline{N} \gamma_5
\bar{D} \Lambda_c + \mbox{ H.c.},
\nonumber \\
\mathcal{L}_{D^*N\Lambda_c} &=& -g_{D^*N\Lambda_c}^{} \left( \overline{N}
\gamma_\mu \bar{D}^{*\mu} - \frac{\kappa_{D^*N\Lambda_c}^{}}{2M_N}
\sigma_{\mu\nu} \partial^\nu \bar{D}^{*\mu} \right) \Lambda_c
\nonumber \\ && \mbox{}
+ \mbox{ H.c.},
\label{eq:Lag}
\end{eqnarray}
where $\psi_\mu$ is the $J/\psi$ vector-meson field and $D$ is the
iso-doublet $D$ meson field, $D = (D^0, D^+)$ and $\bar{D} =
(\bar{D}^0, D^-)^T$. The iso-doublet $D^*$ vector-meson field is
defined in the similar way. In order to determine the coupling
constants, several methods have been suggested and these include
quark models using the heavy quark effective theory approach
\cite{DNP03}, QCD sum rules \cite{MNNR02,RMNN03,BCNN04}, and the
SU(4) symmetry. In this Section, we discuss the coupling constants
in the effective Lagrangians in some detail.

\subsection{\boldmath Vector-meson dominance and the $J/\psi$ couplings}

For the $J/\psi$ couplings we use vector-meson dominance (VMD) as in
Refs.~\cite{MM98a,LK00,OSL01}. In VMD, the photon couples to a
hadron through intermediate vector mesons so that
\begin{equation}
\left. \langle H | J^\mu_{\rm em} | H \rangle = \sum_V
\frac{1}{M_V^2 - p^2} \langle 0 | J^\mu_{\rm em} | V \rangle \langle
H | V | H \rangle \right|_{p^2 \to 0},
\end{equation}
where the sum runs over vector meson states $V$ and the
current-field identity gives $\langle 0 | J^\mu_{\rm em} | V \rangle
= - (M_V^2/f_V^{}) \varepsilon^\mu_V$ with the vector meson
polarization vector $\varepsilon^\mu_V$. The parameter $f_V^{}$ can
be obtained from $\Gamma(V \to e^+e^-) = 4\pi\alpha_{\rm em}^2 M_V/
(3 f_V^2)$. The most recent compilation of the data gives $f_\rho =
4.95$, $f_\omega = 17.10$, $f_\phi = 13.39$, $f_\psi = 11.16$ so
that $f_\rho:f_\omega:f_\phi:f_\psi = 1: 3.45 : 2.70 : 2.25$, while
the SU(4) symmetry gives $f_\rho:f_\omega:f_\phi:f_\psi = 1: 3:
3/\sqrt2 : 3/2\sqrt2 \approx 1:3:2.12:1.06$. This evidently shows
the aspect of the badly broken flavor SU(4) symmetry. Application to
the $\Upsilon$ meson makes the symmetry relation worse as we have
$f_\Upsilon/f_\rho \approx 8$ from the data, while the SU(5)
symmetry implies $f_\Upsilon/f_\rho = 3/\sqrt2$. For the
intermediate state, we consider $V=\rho,\omega,\phi,J/\psi$ by
expecting either that the higher vector meson contributions are
suppressed or that there is a strong cancellation among them,
especially in the charm sector \cite{DNP03}. Applying VMD to the
$D$-meson iso-doublet $(D^0, D^+)$, we then have the following
relations among the coupled equations \cite{LK00}:
\begin{equation}
\left( \begin{array}{c} 0 \\ 1 \end{array} \right) =
\frac{g_{\rho DD}^{}}{f_\rho}
\left( \begin{array}{c} 1 \\ -1 \end{array} \right) +
\frac{g_{\omega DD}^{}}{f_\omega}
\left( \begin{array}{c} 1 \\ 1 \end{array} \right) +
\frac{g_{\psi DD}^{}}{f_\psi}
\left( \begin{array}{c} 1 \\ 1 \end{array} \right),
\end{equation}
which can be solved by using the SU(2) symmetry relations,
$g_{\omega DD}^{} = g_{\rho DD}^{}$ and $f_\omega = 3f_\rho$.
Thus, we have
\begin{equation}
\frac23 = \frac{g_{\psi DD}^{}}{f_\psi}.
\end{equation}
This means that the photon sees the charm quark charge through the
intermediate $J/\psi$ vector-meson.%
\footnote{If we apply VMD to the kaon iso-doublet, we have $-1/3 =
g_{\phi KK}^{}/f_\phi$, which gives $g_{\phi KK}^{} = -4.46$. This
should be compared with $|g_{\phi KK}^{}| = 4.49$ determined from
the experimental data for $\Gamma(\phi \to K \bar{K})$.} By applying
to the $D^*$ iso-doublet, we get the relation, $g_{\psi DD}^{} =
g_{\psi D^*D^*}^{}$. In the QCD sum-rule calculations of
Refs.~\cite{MNNR02,BCNN04}, this relation holds within $20 \sim
30$\%.

For the $J/\psi D^* D$ coupling, we use the relation of the heavy
quark mass limit \cite{DNP03},
\begin{equation}
g_{\psi D^*D}^{} = g_{\psi DD} / \tilde M_{D},
\end{equation}
where $\tilde M_{D}$ is the (average) mass scale of the $D$/$D^*$
mesons. This leads to the coupling constants
\begin{equation}
g_{\psi DD}^{} = g_{\psi D^*D^*}^{} = 7.44, \qquad
g_{\psi D^*D}^{} = 3.84 \mbox{ GeV}^{-1}.
\end{equation}

The coupling constant $g_{\psi\Lambda_c\Lambda_c}^{}$ can also be
estimated through VMD. As we have seen before, the photon sees the
charm quark charge through the $J/\psi$. If we apply this to
$\Lambda_c^+$, then we have
\begin{equation}
g_{\psi\Lambda_c\Lambda_c}^{} = g_{\psi DD}^{} = g_{\psi D^*D^*}^{},
\label{eq:VDM-rel}
\end{equation}
that is, the universality of the $J/\psi$ coupling. This is also
closely related to the charm vector-current conservation as will be
shown below. In Ref.~\cite{LKL02}, $g_{\psi\Lambda_c\Lambda_c}^{}$
was estimated from the SU(4) relation assuming that the $J/\psi$
belongs to the {\bf 15}-multiplet, which gives
$g_{\psi\Lambda_c\Lambda_c}^{} = -1.4$. But, with this assumption,
the $J/\psi$ contains significant light quark components and, as a
result, $g_{\psi\Lambda_c\Lambda_c}^{}$ is underestimated by a
factor of $5$ compared with our estimate.

The tensor coupling constant $\kappa_{\psi\Lambda_c\Lambda_c}^{}$
can also be estimated by using VMD with the anomalous magnetic
moment of $\Lambda_c$. The magnetic moment of $\Lambda_c$ has not
been measured, but the quark model predicts $\mu(\Lambda_c)
\approx 0.37$ \cite{Lic77}, and this gives a predicted anomalous
magnetic moment $\kappa_{\Lambda_c}^{} = -0.63$ for $\Lambda_c$ .
Since the light $u$, $d$ quarks form a spin-$0$ state in
$\Lambda_c$, the $\Lambda_c$ magnetic moment is solely determined by
the charm quark. Therefore, VMD gives
\begin{equation}
\kappa_{\Lambda_c}^{} =
\frac{g_{\psi\Lambda_c\Lambda_c}^{}\kappa_{\psi\Lambda_c\Lambda_c}}{f_\psi},
\end{equation}
which leads to $\kappa_{\psi\Lambda_c\Lambda_c}^{} \approx -0.94$.

\subsection{\boldmath $D$ and $D^*$ meson couplings}

In flavor SU(4), mesons are in a {\bf 15}-multiplet and baryons are
in a {\bf 20}-multiplet, which correspond to the meson octet and
baryon octet of SU(3), respectively. Since ${\bf 15} \otimes
\overline{\bf 20} = {\bf 140} \oplus \overline{\bf 60} \oplus
\overline{\bf 36} \oplus {\bf 20}' \oplus {\bf 20} \oplus {\bf 20}
\oplus \overline{\bf 4}$, there are two couplings for the
meson-baryon-baryon interactions as in the case of SU(3), and they
can be related to the SU(3) coupling constants $D$ and $F$. This
gives the SU(4) symmetry relations, $g_{DN\Lambda_c}^{} =
g_{KN\Lambda}^{}$ and $g_{D^*N\Lambda_c}^{} = g_{K^*N\Lambda}^{}$,
of Ref.~\cite{LKL02}. The empirical values of Ref.~\cite{SR99} for
strange hadrons then give
\begin{equation}
g_{DN\Lambda_c}^{} = -13.2, \quad g_{D^*N\Lambda_c}^{} = -4.3.
\label{eq:Nij}
\end{equation}
These values are quite different from the QCD sum-rule predictions
of Ref.~\cite{DNN00}
\begin{equation}
|g_{DN\Lambda_c}^{}| = 7.9, \quad |g_{D^*N\Lambda_c}^{}| = 7.5.
\label{eq:qcdsr}
\end{equation}
This difference can affect the value of the ratio $R_{D/D^*}$ as will be
discussed below.

For the tensor coupling constant $\kappa_{D^*N\Lambda_c}^{}$, there is no
theoretical prediction for its value.
If we assume the SU(4) relation again, we have \cite{SR99}
\begin{equation}
\kappa_{D^*N\Lambda_c}^{} = 2.65.
\label{eq:Nij2}
\end{equation}
However, it should be mentioned that the SU(4) symmetry breaking
effects can significantly alter the values of the coupling constants
given in Eqs.~(\ref{eq:Nij}) and (\ref{eq:Nij2}). Therefore, in this
work, we investigate the role of the $DN\Lambda_c$ and
$D^*N\Lambda_c$ interactions by varying their coupling constants.

The non-vanishing tensor coupling for the $D^*N\Lambda_c$
interaction also causes the four-point interaction that is shown in
Fig.~\ref{fig:diag}(c). This term can be obtained by gauging the
tensor interaction and it reads
\begin{eqnarray}
\mathcal{L}_{\psi D^*N\Lambda_c} &=& -i \frac{g_{\psi}^{}}{2M_N}
g_{D^* N\Lambda_c}^{} \kappa_{D^*N\Lambda_c}^{} \bar{N}
\sigma_{\mu\nu}^{} \bar{D}^{*\mu} \psi^\nu \Lambda_c
\nonumber \\ && \mbox{}
+ \mbox{ H.c.},
\end{eqnarray}
where $g_\psi^{}$ is the gauge coupling constant. As we shall see
below, $g_\psi^{}$ can be related to the universal $J/\psi$ coupling
constant by charm vector-current conservation.

\subsection{Production amplitudes and form factors}

The production amplitudes can be written as
\begin{eqnarray}
\mathcal{M}(J/\psi + N \to \bar{D} + \Lambda_c) &=& \mathcal{M}^\mu_D
\varepsilon_\mu(\psi),
\nonumber \\
\mathcal{M}(J/\psi + N \to \bar{D}^* + \Lambda_c) &=&
\varepsilon^*_\nu (D^*) \mathcal{M}^{\mu\nu}_{D^*} \varepsilon_\mu(\psi),
\end{eqnarray}
with
\begin{eqnarray}
\mathcal{M}^\mu_D &=& \mathcal{M}^\mu_t + \mathcal{M}^\mu_u +
\mathcal{M}^\mu_{\rm an},
\nonumber \\
\mathcal{M}^{\mu\nu}_{D^*} &=& \mathcal{M}^{\mu\nu}_t +
\mathcal{M}^{\mu\nu}_u + \mathcal{M}^{\mu\nu}_{\rm an} +
\mathcal{M}_c^{\mu\nu},
\end{eqnarray}
where $\mathcal{M}_t^{}$ is the $t$-channel amplitude,
$\mathcal{M}_u^{}$ is the $u$-channel amplitude, $\mathcal{M}_{\rm
an}^{}$ is the $t$-channel amplitude including the anomalous $J/\psi
D^*D$ interaction, and $\mathcal{M}_c^{}$ is from the contact term.
The amplitudes can be obtained straightforwardly from the
interaction Lagrangians, e.g., as in Ref.~\cite{LKL02}, and will not
be given here.

Now we impose the conservation condition of the charm vector-current
to the production amplitudes, i.e., ${p_\psi^{}}_\mu
\mathcal{M}_{D}^{\mu} = 0$ and ${p_\psi^{}}_\mu
\mathcal{M}_{D^*}^{\mu\nu} = 0$, where $p_\psi^{}$ is the
four-momentum of the $J/\psi$. The anomalous terms and the
$\kappa_{\psi\Lambda_c\Lambda_c}^{}$ terms already satisfy this
condition separately, and we have
\begin{eqnarray}
&&
{p_\psi^{}}_\mu \mathcal{M}_D^{\mu} =
i g_{DN\Lambda_c}^{} \left( g_{\psi
DD}^{} - g_{\psi \Lambda_c\Lambda_c}^{} \right) \gamma_5^{},
\nonumber \\ &&
{p_\psi^{}}_\mu \mathcal{M}_{D^*}^{\mu\nu} =  g_{D^*N\Lambda_c}^{}
\left( g_{\psi\Lambda_c\Lambda_c}^{} - g_{\psi D^* D^*}^{} \right)
\gamma^\nu
\nonumber \\ && \mbox{}
+ \frac{g_{D^*N\Lambda_c}^{}\kappa_{D^*N\Lambda_c}^{}}{2M_N}
\sigma^{\nu\lambda}
\bigl\{ p_{D^*\lambda}^{} (g_{\psi D^*D^*}^{} -
g_{\psi\Lambda_c\Lambda_c}^{})
\nonumber \\ && \mbox{} \qquad
+p_{\psi\lambda}^{} (g_\psi^{} - g_{\psi D^*D^*}^{})\bigr\},
\end{eqnarray}
where $p_{D^*}^{}$ is the four-momentum of the produced $D^*$ meson.
Thus, vector-current conservation leads to
\begin{equation}
g_\psi^{} = g_{\psi\Lambda_c\Lambda_c}^{} = g_{\psi DD}^{}
= g_{\psi D^*D^*}^{}.
\end{equation}
Therefore, one can verify that the charm vector-current conservation
leads to the VMD relation (\ref{eq:VDM-rel}) for the coupling constants
and fixes the gauge coupling constant $g_\psi^{}$.

\begin{figure*}[t]
\centering
\epsfig{file=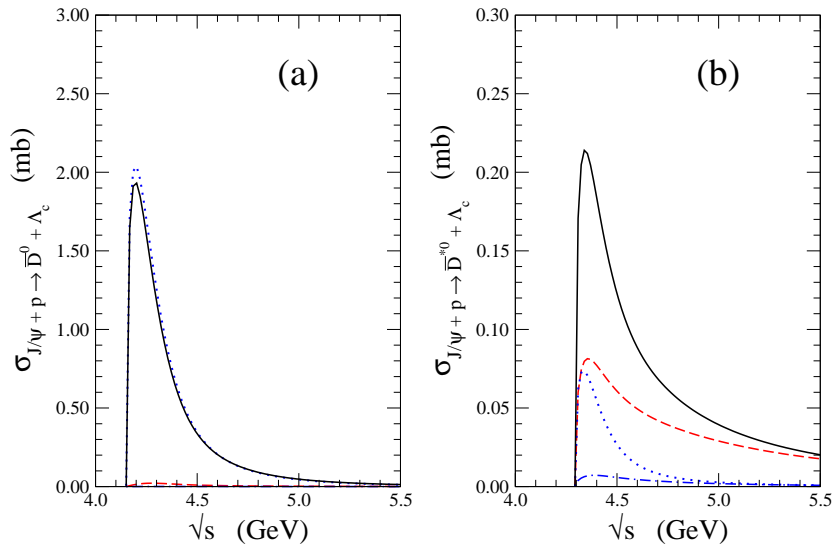, width=0.5\textwidth, angle=-90}
\caption{(Color online)
Total cross section for (a) $J/\psi + p \to \bar{D}^{0} +
\Lambda_c$ and (b) $J/\psi + p \to \bar{D}^{*0} + \Lambda_c$ when the
tensor interactions are turned off, $\kappa_{\psi\Lambda_c\Lambda_c}^{}
= \kappa_{D^*N\Lambda_c}^{}=0$.
The dashed, dotted, and dot-dashed lines are the contributions from
the $t$-channel, $u$-channel, and anomalous terms, respectively, and the
solid lines are their sums.}
\label{fig:xs}
\end{figure*}

Because of the finite size of hadrons, it is required to include
form factors in effective Lagrangian approaches, which are functions
of the momentum of exchanged (or off-shell) particles. The form
factors may be calculated from more microscopic theories
\cite{CDNN05,DNN00}, but here we employ a simple phenomenological
form \cite{PJ91}
\begin{equation}
F(p_{\rm ex}^2) = \left( \frac{n\Lambda^4}{n\Lambda^4 + (p_{\rm ex}^2 -
M_{\rm ex}^2)^2} \right)^n,
\end{equation}
where $p_{\rm ex}^{}$ and $M_{\rm ex}^{}$ are the four-momentum and
mass of the exchanged particle, respectively. Therefore, when the
exchanged particle is on its mass-shell, it has the correct
normalization $F(p_{\rm ex}^2 = M_{\rm ex}^2) = 1$ and, as $n \to
\infty$, $F(p^2)$ becomes a Gaussian of $(p^2-M_{\rm ex}^2)$ with a
width of $\Lambda^2$. In this work, we take the limit $n \to
\infty$.

However, employing such form factors violates the current
conservation condition. There is no unique way to restore current
conservation with form factors and in this work we follow the
prescription of Ref.~\cite{DW01a}, namely, current conservation is
recovered by introducing the contact diagram of
Fig.~\ref{fig:diag}(c), which is, in practical calculation,
equivalent to replace the form factors by a universal one in the
form of
\begin{equation}
1 - [1-F(s)][1-F(u)].
\label{eq:cff}
\end{equation}
This form factor is also employed for $\bar{D}^*\Lambda_c$ production in
the presence of the $D^*N\Lambda_c$ tensor interaction.

\section{Results}

We first discuss our results without the tensor coupling terms,
i.e., by setting $\kappa_{\psi\Lambda_c\Lambda_c}^{} =
\kappa_{D^*N\Lambda_c}^{} = 0$ for a comparison with the results of
Ref.~\cite{LKL02}. Shown in Fig.~\ref{fig:xs} are our results on the
total cross sections for $J/\psi + p \to \bar{D}^{(*)0} + \Lambda_c$
obtained with the couplings of Eq.~(\ref{eq:Nij}). The results
depend on the cutoff $\Lambda$, and $\Lambda=1.8$ GeV is used for
this calculation. The dependence of our results on $\Lambda$ will be
discussed later. The dashed, dotted, and dot-dashed lines are the
contributions from the $t$-channel, $u$-channel, and anomalous
terms, respectively, and the solid lines are their sums. There are
several comments in comparing with the results of Ref.~\cite{LKL02}.
We first verify the conclusion of Ref.~\cite{LKL02} that the
anomalous interaction terms give small contributions in both
reactions. However, there are several crucial differences. Since the
VMD and current conservation condition require much larger
$g_{\psi\Lambda_c\Lambda_c}^{}$ coupling constant, this enhances the
contribution from the $u$-channel diagram in both reactions. As a
result, the $J/\psi + p \to \bar{D}^{0} + \Lambda_c$ reaction is
dominated by the $u$-channel diagram, and the $J/\psi + p \to
\bar{D}^{*0} + \Lambda_c$ has comparable contributions from both the
$t$- and $u$-channel diagrams. Furthermore, this makes the cross
section ratio $R_{D/D^*}$ to be larger than $1$, which is opposite
to the result of Ref.~\cite{LKL02}. In addition, our predictions on
the energy dependence of the cross sections show more rapid decrease
of the cross sections at larger energies than that of
Ref.~\cite{LKL02}. Although this is partly due to the Gaussian form
factor adopted in this model, it is the current conserved form of
the form factors (\ref{eq:cff}) that is mainly responsible for this
energy dependence of the cross sections. Taking into account all
these effects, we found that $R_{D/D^*} \simeq 10$ with our
parameters, and the peak value of the cross sections for $\bar{D}^0
\Lambda_c$ final state reaction is close to $2$ mb.

\begin{figure*}[t]
\centering \epsfig{file=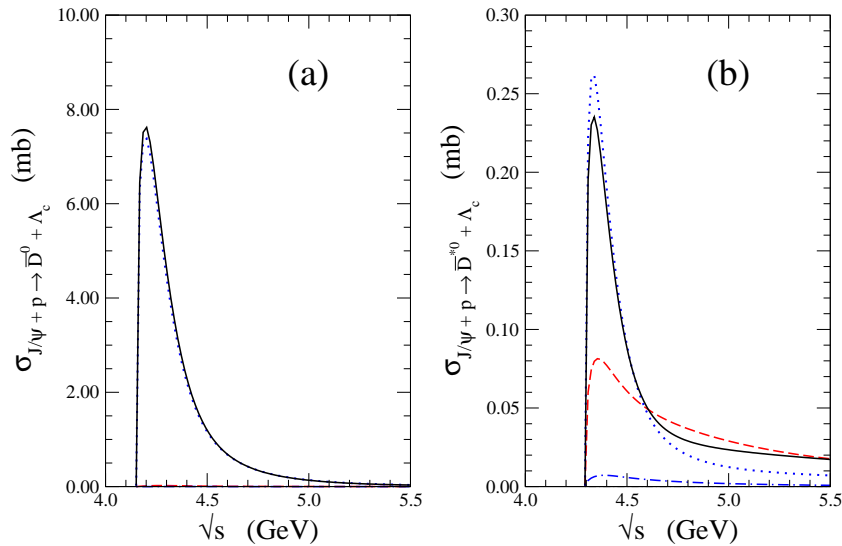, width=0.5\textwidth, angle=-90}
\caption{(Color online)
Total cross section for (a) $J/\psi + p \to \bar{D}^{0} +
\Lambda_c$ and (b) $J/\psi + p \to \bar{D}^{*0} + \Lambda_c$ with
$\kappa_{\psi\Lambda_c\Lambda_c}^{} = -0.94$ and
$\kappa_{D^*N\Lambda_c}^{}=0$. Notations are the same as in
Fig.~\ref{fig:xs}.} \label{fig:xs-psi}
\end{figure*}

We also found that the tensor interactions,
$\kappa_{\psi\Lambda_c\Lambda_c}^{}$ term and
$\kappa_{D^*N\Lambda_c}^{}$ term, can give nontrivial contributions
to the cross sections but play a different role. To see the role of
these terms, we first present the results with
$\kappa_{\psi\Lambda_c\Lambda_c}^{} = -0.94$ but with
$\kappa_{D^*N\Lambda_c}^{} = 0$ in Fig.~\ref{fig:xs-psi}, while
keeping the other parameters as in the case of Fig.~\ref{fig:xs}. In
this case, the $u$-channel contributions are further enhanced.
Furthermore, because of the large contributions from the
$\kappa_{\psi\Lambda_c\Lambda_c}^{}$ term, the $t$-channel and
$u$-channel diagrams interfere constructively in $\bar{D}\Lambda_c$
production and destructively in $\bar{D}^*\Lambda_c$ production,
which is opposite to the results of Fig.~\ref{fig:xs}. Consequently,
the $J/\psi + N \to \bar{D}\Lambda_c$ cross sections are enhanced,
but the $J/\psi + N \to \bar{D}^*\Lambda_c$ cross sections are not
changed so much. This leads to the increase of the cross section
ratio and we have $R_{D/D^*} \simeq 30$, which is close to the
prediction of the quark-interchange model of Ref.~\cite{HBBS07}.

\begin{figure*}[t]
\centering \epsfig{file=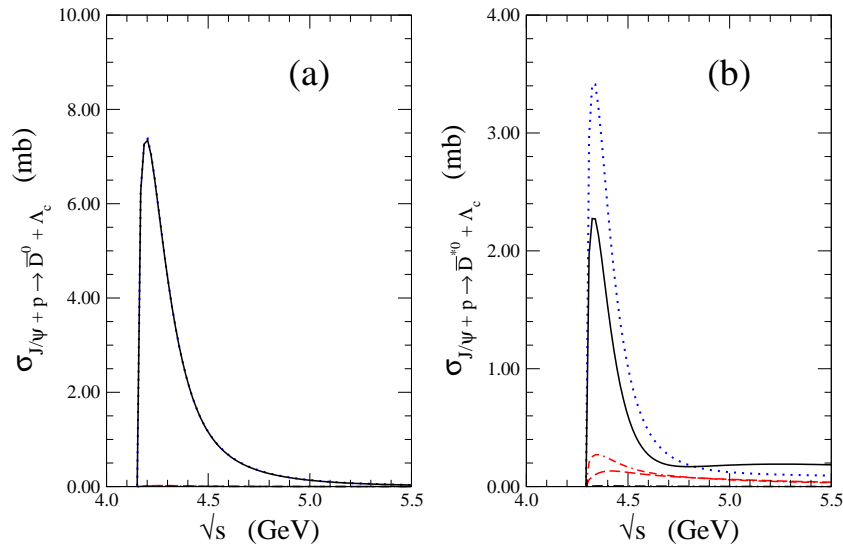, width=0.5\textwidth, angle=-90}
\caption{(Color online)
Total cross section for (a) $J/\psi + p \to \bar{D}^{0} +
\Lambda_c$ and (b) $J/\psi + p \to \bar{D}^{*0} + \Lambda_c$ with
$\kappa_{\psi\Lambda_c\Lambda_c}^{} = -0.94$ and
$\kappa_{D^*N\Lambda_c}^{}=2.65$. In (a), the $u$-channel diagram
dominates and the contributions from the other diagrams are
suppressed and cannot be seen. In (b), the dot-dash-dashed line is
the contribution from the contact term. Other notations are the same
as in Fig.~\ref{fig:xs}.} \label{fig:xs-dstar}
\end{figure*}

To verify the role of the $\kappa_{D^*N\Lambda_c}^{}$ tensor term,
we use the value of Eq.~(\ref{eq:Nij2}) since there is no
theoretical prediction for this coupling constant. Other parameters
are the same as in the case of Fig.~\ref{fig:xs-psi} and the results
are given in Fig.~\ref{fig:xs-dstar}. This shows that the effect of
the $\kappa_{D^*N\Lambda_c}^{}$ term in $J/\psi + N \to \bar{D} +
\Lambda_c$ is negligible, which is expected since it contributes to
the suppressed $t$-channel anomalous term only. However, this tensor
interaction can change noticeably the cross sections for the $J/\psi
+ N \to \bar{D}^* + \Lambda_c$ reaction. This is because the tensor
term enters both in the $t$-channel and in the $u$-channel diagrams.
Furthermore, this term requires the presence of the contact term. As
a result, the $\kappa_{D^*N\Lambda_c}^{}$ term enhances the
$\bar{D}^*\Lambda_c$ production cross sections and leads to a
smaller value of the ratio $R_{D/D^*}$, and $R_{D/D^*} \simeq 3$ is
observed in Fig.~\ref{fig:xs-dstar}.

The cross sections and their ratio $R_{D/D^*}$ depend on the
coupling constants $g_{DN\Lambda_c}^{}$ and $g_{D^*N\Lambda_c}^{}$
which are not well understood yet. In order to see this dependence,
we use the coupling constants (\ref{eq:qcdsr}) predicted by the QCD
sum-rule calculation of Ref.~\cite{DNN00}. This gives very different
values of the cross section ratio, and instead of $R_{D/D^*} \sim
30$ and $\sim 3$, (Figs.~\ref{fig:xs-psi} and \ref{fig:xs-dstar}) we
have $R_{D/D^*} \sim 8$ and $\sim 0.7$, respectively. This also
changes the corresponding maximum values of the $J/\psi + p \to
\bar{D}^0 + \Lambda_c$ cross sections and they are $\sim 1.6$ mb and
$\sim 1.4$ mb for the two cases, respectively, as shown in
Table~\ref{tab1}.

As was mentioned before, the cross sections also depend on the
cutoff parameter $\Lambda$, and we have used $\Lambda = 1.8$ GeV. In
order to see the dependence of our results on the form factor, we
repeat the calculation for three different values of the cutoff,
i.e., $\Lambda=1.5$ GeV, $1.8$ GeV, and $2.1$ GeV. Shown in
Table~\ref{tab1} are the peak values of the $J/\psi + N \to \bar{D}
+ \Lambda_c$ cross sections and the ratio $R_{D/D^*}$. We found that
the cross sections for the $J/\psi + N \to \bar{D}^* + \Lambda_c$
reaction are more sensitive to the cutoff than those for the $J/\psi
+ N \to \bar{D} + \Lambda_c$ reaction. Thus the ratio $R_{D/D^*}$
decreases as the cutoff parameter $\Lambda$ increases. However, we
found that the $\kappa_{D^*N\Lambda_c}^{}$ term suppresses
$R_{D/D^*}$ regardless of the cutoff parameter value.

\begin{table*}[t]
\centering
\begin{tabular}{c|c|cc|cc}  \hline\hline
$\Lambda$ (GeV) & &
\multicolumn{2}{c|}{With the couplings in Eq.~(\ref{eq:Nij})} &
\multicolumn{2}{c}{With the couplings in Eq.~(\ref{eq:qcdsr})}
\\ \cline{3-6}
 & & (I) & (II) & (I) & (II) \\ \hline
$1.5$ & $\sigma(J/\psi+N\to\bar{D}+\Lambda_c)$ &
$0.86$ mb & $0.86$ mb & $0.3$ mb & $0.3$ mb \\
                    & $R_{D/D^*}$ & $\sim 225$ & $\sim 30$
& $\sim 38$ & $\sim 3$ \\ \hline
$1.8$ & $\sigma(J/\psi+N\to\bar{D}+\Lambda_c)$ &
$7.5 $ mb & $7.5 $ mb & $1.6$ mb & $1.4$ mb \\
                    & $R_{D/D^*}$ & $\sim 30 $ & $\sim 3 $
& $\sim 8$ & $\sim 0.7$ \\ \hline
$2.1$ & $\sigma(J/\psi+N\to\bar{D}+\Lambda_c)$ &
$20.5$ mb & $19.5$ mb & $6.6$ mb & $6.1$ mb \\
                    & $R_{D/D^*}$ & $\sim 12 $ & $\sim 1.2 $
& $\sim 1.4$ & $\sim 0.12$ \\ \hline\hline
\end{tabular}
\caption{The peak values of the cross section for
$J/\psi+N\to\bar{D}+\Lambda_c)$ and the ratio $R_{D/D^*}$ for
different choices of the $D/D^*$ coupling constants and for three
different values of the cutoff parameter $\Lambda$. (I) is for
$\kappa_{D^*N\Lambda_c}^{}=0$, and (II) is for
$\kappa_{D^*N\Lambda_c}^{}=2.65$. We use
$\kappa_{\psi\Lambda_c\Lambda_c}^{} = -0.94$ for the both cases.}
\label{tab1}
\end{table*}

\section{Conclusion}

In this work, we have reanalyzed and improved the meson-exchange
model for $J/\psi$-nucleon reaction of Ref.~\cite{LKL02}. We found
that vector-meson dominance and charm vector-current conservation
lead to the universality of the $J/\psi$ meson couplings, which can
drastically change the ratio $R_{D/D^*}$ of the cross sections of
$J/\psi + N \to \bar{D} + \Lambda_c$ and $J/\psi + N \to \bar{D}^* +
\Lambda_c$. It is also found that this ratio is sensitive to the
relative strengths of $g_{DN\Lambda_c}^{}$ and
$g_{D^*N\Lambda_c}^{}$ as well as to the tensor coupling terms in
the $J/\psi\Lambda_c\Lambda_c$ and $D^*N\Lambda_c$ interactions.
We found that the VMD and
vector-current conservation lead to a large value of $R_{D/D^*}$.
This value can be further enhanced by the $J/\psi\Lambda_c\Lambda_c$
tensor interaction. But the $D^*N\Lambda_c$ tensor interaction has
the opposite role by decreasing $R_{D/D^*}$.

To match the quark-interchange model predictions of
Ref.~\cite{HBBS07} with those from our effective Lagrangian approach
leads us to conclude that $g_{DN\Lambda_c}^{}$ must be larger than
$g_{D^*N\Lambda_c}^{}$ and $\kappa_{D^*N\Lambda}$ must be small. The
first condition contradicts with the QCD sum-rule predictions of
Ref.~\cite{DNN00} that prefers a similar strength for the two
couplings. Instead, the SU(4) symmetry relations satisfy this
condition. However, SU(4) symmetry gives a large value for
$\kappa_{D^*N\Lambda}$, and thus does not fulfill the second
condition. Since SU(4) symmetry must be broken by the heavy charm
quark mass, it would be interesting to see how badly the SU(4)
symmetry relations for $g_{DN\Lambda_c}^{}$, $g_{D^*N\Lambda_c}^{}$,
and $\kappa_{D^*N\Lambda_c}^{}$ are broken. Therefore, more rigorous
studies on these couplings are required, which will eventually help
to reconcile the predictions of the quark-interchange model and of the
meson-exchange model. Nevertheless, the constraints used in this
work, VMD and charm vector-current conservation, are found to have a
nontrivial role to fill the gap between the quark-interchange model
and meson-exchange model predictions to some extent.

\acknowledgments

This work was supported in part by the US National Science Foundation under
Grant No. PHY-0457265, and the Welch Foundation under Grant No. A-1358.


\begin{thebibliography}{10}

\bibitem{MS86}
T.~Matsui and H.~Satz,
\newblock Phys. Lett. B {\bf 178}, 416 (1986).

\bibitem{FLP88-GH88-GGJ88-VPKH88}
J.~Ft\'{a}\v{c}nik, P.~Lichard, and J.~Pi\v{s}\'{u}t,
\newblock Phys. Lett. B {\bf 207}, 194 (1988);
C.~Gerschel and J.~H{\"u}fner,
\newblock {\it ibid.\/} {\bf 207}, 253 (1988);
S.~Gavin, M.~Gyulassy, and A.~Jackson,
\newblock {\it ibid.\/} {\bf 207}, 257 (1988);
R.~Vogt, M.~Prakash, P.~Koch, and T.~H. Hansson,
\newblock {\it ibid.\/} {\bf 207}, 263 (1988).

\bibitem{HK98-RSZ00}
J.~H{\"u}fner and B.~Z. Kopeliovich,
\newblock Phys. Lett. B {\bf 426}, 154 (1998);
K.~Redlich, H.~Satz, and G.~M. Zinovjev,
\newblock Eur. Phys. J. C {\bf 17}, 461 (2000).

\bibitem{KLNS97}
D.~Kharzeev, C.~Louren\c{c}o, M.~Nardi, and H.~Satz,
\newblock Z. Phys. C {\bf 74}, 307 (1997).

\bibitem{AT06}
F.~Arleo and V.-N. Tram,
hep-ph/0612043.

\bibitem{pqcd}
G.~Bhanot and M.~E. Peskin,
\newblock Nucl. Phys. {\bf B156}, 391 (1979);
D.~Kharzeev and H.~Satz,
\newblock Phys. Lett. B {\bf 334}, 155 (1994);
Y.~Oh, S.~Kim, and S.~H. Lee,
\newblock Phys. Rev. C {\bf 65}, 067901 (2002);
S.~H. Lee and Y.~Oh,
\newblock J. Phys. G {\bf 28}, 1903 (2002);
F.~Arleo, P.-B. Gossiaux, T.~Gousset, and J.~Aichelin,
\newblock Phys. Rev. D {\bf 65}, 014005 (2002);
T.~Song and S.~H. Lee,
\newblock {\it ibid.\/} {\bf 72}, 034002 (2005).

\bibitem{qcdsr}
D.~Kharzeev, H.~Satz, A.~Syamtomov, and G.~Zinovjev,
\newblock Phys. Lett. B {\bf 389}, 595 (1996);
F.~S. Navarra, M.~Nielsen, R.~S. Marques~de Carvalho, and G.~Krein,
\newblock {\it ibid.\/} {\bf 529}, 87 (2002);
F.~O. Dur{\~a}es, S.~H. Lee, F.~S. Navarra, and M.~Nielsen,
\newblock {\it ibid.\/} {\bf 564}, 97 (2003);
F.~O. Dur{\~a}es, H.~Kim, S.~H. Lee, F.~S. Navarra, and M.~Nielsen,
\newblock Phys. Rev. C {\bf 68}, 035208 (2003).

\bibitem{Haglin00-STT01}
K.~L. Haglin,
\newblock Phys. Rev. C {\bf 61}, 031902 (2000);
A.~Sibirtsev, K.~Tsushima, and A.~W. Thomas,
\newblock {\it ibid.\/} {\bf 63}, 044906 (2001);
K.~L. Haglin and C.~Gale,
\newblock {\it ibid.\/} {\bf 63}, 065201 (2001);
A.~Bourque, C.~Gale, and K.~L. Haglin,
\newblock {\it ibid.\/} {\bf 70}, 055203 (2004).

\bibitem{OSL01}
Y.~Oh, T.~Song, and S.~H. Lee,
\newblock Phys. Rev. C {\bf 63}, 034901 (2001);
Y.~Oh, T.~Song, S.~H. Lee, and C.-Y. Wong,
\newblock J. Korean Phys. Soc. {\bf 43}, 1003 (2003).

\bibitem{MM98a}
S.~G. Matinyan and B.~M{\"u}ller,
\newblock Phys. Rev. C {\bf 58}, 2994 (1998).

\bibitem{LK00}
Z.~Lin and C.~M. Ko,
\newblock Phys. Rev. C {\bf 62}, 034903 (2000).

\bibitem{LKL02}
W.~Liu, C.~M. Ko, and Z.~W. Lin,
\newblock Phys. Rev. C {\bf 65}, 015203 (2002).

\bibitem{IL03-LC06}
A.~\mbox{Yu}. Illarionov and G.~I. Lykasov,
\newblock in {\em Proceedings of International Conference: I.~Ya. Pomeranchuk
  and Physics at the Turn of Centuries}, edited by A.~V. Berkov, N.~B.
  Narozhny, and L.~B. Okun, pp. 305--316, Singapore, 2003, World Scientific,
  hep-ph/0305117;
G.~I. Lykasov and W.~Cassing,
\newblock Nucl. Phys. {\bf A782}, 396c (2007).

\bibitem{qm}
K.~Martins, D.~Blaschke, and E.~Quack,
\newblock Phys. Rev. C {\bf 51}, 2723 (1995);
C.~Y. Wong, E.~S. Swanson, and T.~Barnes,
\newblock {\it ibid.\/} {\bf 62}, 045201 (2000);
\newblock {\bf 65}, 014903 (2002),
\newblock {\bf 66}, 029901(E) (2002);
T.~Barnes, E.~S. Swanson, C.~Y. Wong, and X.~M. Xu,
\newblock {\it ibid.\/} {\bf 68}, 014903 (2003);
M.~A. Ivanov, J.~G. K{\"o}rner, and P.~Santorelli,
\newblock Phys. Rev. D {\bf 70}, 014005 (2004).

\bibitem{HBBS07}
J.~P. Hilbert, N.~Black, T.~Barnes, and E.~S. Swanson, nucl-th/0701087.

\bibitem{YSHH06}
K.~Yokokawa, S.~Sasaki, T.~Hatsuda, and A.~Hayashigaki,
\newblock Phys. Rev. D {\bf 74}, 034504 (2006).

\bibitem{DNNR99-SV05b}
H.~G. Dosch, F.~S. Navarra, M.~Nielsen, and M.~Rueter,
\newblock Phys. Lett. B {\bf 466}, 363 (1999);
A.~Sibirtsev and M.~B. Voloshin,
\newblock Phys. Rev. D {\bf 71}, 076005 (2005).

\bibitem{DNP03}
A.~Deandrea, G.~Nardulli, and A.~D. Polosa,
\newblock Phys. Rev. D {\bf 68}, 034002 (2003).

\bibitem{MNNR02}
R.~D. Matheus, F.~S. Navarra, M.~Nielsen, and R.~Rodrigues~da Silva,
\newblock Phys. Lett. B {\bf 541}, 265 (2002).

\bibitem{RMNN03}
R.~Rodrigues~da Silva, R.~D. Matheus, F.~S. Navarra, and M.~Nielsen,
\newblock Braz. Jour. Phys. {\bf 34}, 236 (2004).

\bibitem{BCNN04}
M.~E. Bracco, M.~Chiapparini, F.~S. Navarra, and M.~Nielsen,
\newblock Phys. Lett. B {\bf 605}, 326 (2005).

\bibitem{Lic77}
D.~B. Lichtenberg,
\newblock Phys. Rev. D {\bf 15}, 345 (1977).

\bibitem{DNN00}
F.~O. Dur{\~a}es, F.~S. Navarra, and M.~Nielsen,
\newblock Phys. Lett. B {\bf 498}, 169 (2001).

\bibitem{SR99}
V.~G.~J. Stoks and \mbox{Th}. A.~Rijken,
\newblock Phys. Rev. C {\bf 59}, 3009 (1999).

\bibitem{CDNN05}
F.~Carvalho, F.~O. Dur{\~a}es, F.~S. Navarra, and M.~Nielsen,
\newblock Phys. Rev. C {\bf 72}, 024902 (2005).

\bibitem{PJ91}
B.~C. Pearce and B.~K. Jennings,
\newblock Nucl. Phys. {\bf A528}, 655 (1991).

\bibitem{DW01a}
R.~M. Davidson and R.~Workman,
\newblock Phys. Rev. C {\bf 63}, 025210 (2001).

\end{thebibliography}
\end{document}